# Microtubule-based active fluids with improved lifetime, temporal stability and miscibility with passive soft materials

Pooja Chandrakar,*[a,g] John Berezney [a], Bezia Lemma,[a,b,g] Bernard Hishamunda,[a] Angela Berry,[a] Kun-Ta Wu,[a,c] Radhika Subramanian,[d] Johnson Chung,[e] Daniel Needleman,[b,f] Jeff Gelles,[e] and Zvonimir Dogic [a,g]

We have developed several distinct model systems of microtubule-based 3D active isotropic fluids and have compared their dynamical and structural properties. The non-equilibrium dynamics of these fluids is powered by three different types of kinesin motors: a processive motor, a non-processive motor, and a motor which is permanently linked to a microtubule backbone. The fourth modification uses specific microtubule crosslinkers to induce bundle formation instead of a non-specific depletant. In comparison to the formerly developed systems, each new active fluid formulation has certain enhanced properties. Firstly, linking motors to the microtubule backbone enhances the fluid lifetime from hours to several days. Furthermore, switching to the non-processive motors significantly increases the temporal stability of the active dynamics, while using specific cross-linkers yields systems that can incorporate other passive soft materials, such as, polymer gels or liquid crystals. These novel developed model systems will significantly aid and improve our ability to quantify diverse phenomena observed in microtubule based active matter.

## I. Introduction

Studies of active matter are focused on elucidating the fundamental laws, which govern the non-equilibrium dynamics that emerges in assemblages of motile interacting entities[1-3]. The Self-organized active matter phenomena span many length scales, ranging from the coordinated movement of emperor penguins while they huddle[4], to cell organization assisting the folding of epithelial tissue[5, 6], and even to the spontaneous flows observed within individual cells[7]. However, many biological realizations of active matter consist of poorly characterized components; making it difficult to compare the experimental results to the microscopically formulated theoretical models. Overcoming these obstacles require the development of novel experimental model systems, where the constituent interactions can be precisely tuned; and consequently, provide a clear pathway to understand the fundamental physical laws, governing the behavior of the self-organized active materials. Promising systems are being built from biological materials such as bacteria[8, 9] and cells[10] while others are being assembled from chemically-fuelled Janus swimmers[11, 12]. Each category has its distinct advantages. Active materials consisting of living bacteria and motile eukaryotic cells are energy efficient; thus, they exhibit long-term dynamics. In comparison, synthetic elements such as Janus swimmers are robustly tuneable. The active systems, based on cytoskeletal building blocks, microtubules (MTs) or actin filaments and associated motor proteins, have multiple attractive attributes. The microscopic dynamics of these systems can be precisely tuned, while energy efficient molecular motors can power long-lived non-equilibrium steady states. These unique features led to the observation of diverse phenomena in cytoskeletal active matter, ranging from the active nematic liquid crystals and the contractile gels, to the collective polar flocks[13-18].

We focus exclusively on the 3D isotropic active fluids, consisting of MTs and molecular motors. In such fluids, the extensile MT bundles undergo repeating cascade of kinesin motor-driven extension, buckling, fracturing and annealing. These dynamics power spontaneous flows that persist for hours[19, 20]. The statistical properties of these flows can be tuned by varying the concentration of motors, ATP, MTs and depletant[21]. Here, we characterize the temporal dynamics of the previously developed active isotropic fluids and its intrinsic dependence on the concentration of the kinesin clusters. We also improve the salient properties of these active fluids in multiple ways. First, instead of using soluble motor clusters we permanently link kinesin onto the MT backbone, which results in the increase of the sample lifetime from a few hours to several days. Second, we power the active fluids by non-processive single-headed kinesin motors instead of the highly processive double-headed motors. This significantly reduces the motor-induced interference and cross-linking, and simultaneously improves the temporal

[a.] The Martin Fisher School of Physics, Brandeis University, Waltham, Massachusetts 02454, USA
[b.] School of Engineering and Applied Sciences, Harvard University, Cambridge, Massachusetts 02138, USA
[c.] Department of Physics, Worcester Polytechnic Institute, 100 Institute Road, Worcester, Massachusetts 01609, USA
[d.] Department of Genetics, HMS and Department of Molecular Biology, Massachusetts General Hospital, Boston, Massachusetts 02114, USA
[e.] Department of Biochemistry, Brandeis University, Waltham, Massachusetts 02454, USA
[f.] Department of Molecular and Cellular Biology, and FAS Center for Systems Biology, Harvard University, Cambridge, Massachusetts 02138, USA
[g.] Department of Physics, University of California, Santa Barbara, California 93106, USA.

stability of the non-equilibrium dynamics. Third, we demonstrate a depletant-free formulation of isotropic active fluids. We use non-motor MT cross-linkers to bundle filaments; thus, increasing the compatibility of MT based active matter with other soft materials. Our findings demonstrate that the dynamics of the extensile isotropic active fluids is robust against changes in the microscopic building blocks. Furthermore, the same structural modifications, developed for active isotropic fluids, can also be extended to improve the properties of MT-based active nematics[22-24].

## II. Building blocks of MT-based active fluids

In this section, we describe the biochemical details of the basic building blocks of the active fluids studied here, and the rationale for developing new model systems. Previously developed active isotropic fluids were assembled from three components: MTs, kinesin clusters and a bundling agent. MTs are rigid hollow cylindrical polar filaments with distinct plus and minus ends that are polymerized from tubulin dimers. In order to suppress dynamic instability, MTs were stabilized by non-hydrolysable GTP analogue, GMPCPP (Guanosine 5'-(α,β-methylenetriphosphate))[25].

The second basic component of the active fluids was kinesin-1 molecular motors that transform chemical energy from the ATP hydrolysis to mechanical motion. Each ATP molecule powers a single 8 nm step of a kinesin dimer towards the MT plus end[26]. In previously developed fluids, the kinesin motors were bound into multi-motor clusters that can simultaneously attach to multiple MTs, thereby inducing their relative sliding and generating active stresses. The third component was the non-adsorbing polymer, which induces attractive forces among MTs through the depletion interaction, leading to their bundling; while still allowing the filaments to slide past each other[27-29]. In the presence of kinesin and a depletant, inter-filament sliding occurs between two anti-parallel MTs[19]. When compared to isotropic suspensions, the bundling mechanism greatly enhances the efficiency of inter-filament sliding, since the motor clusters are more likely to bind to multiple filaments in a bundle configuration. The conventional active isotropic fluids have previously been explored quite extensively; however, their lifetimes and temporal stability have not been quantified[21].

Conventional formulation of active fluids uses motile fragments of biotinylated kinesin-1 motors, which are bound into multi-motor clusters with tetrameric streptavidin (**Fig. 1a**). K401 is a kinesin-1 fragment which spontaneously dimerizes. It is a highly processive motor, which takes ~100 consecutives 8 nm steps, before dissociating from the MT[30, 31]. During this entire duration the motor remains firmly linked to the filament. Highly processive motors can negatively interfere with each other. For example, motility assays showed that increasing motor density leads to a marked slowdown in the MT translocation velocity[32]. This is explicable in terms of the duration of the actual mechanical step, which is a few orders of magnitude smaller than the dwell time between individual steps, during which the motor is firmly attached to the MT. If motors in an ensemble are mechanically coupled, it is expected that each stepping motor experiences a hindering load; since it has to work against other motors during an actual step, which are most likely in the pausing state. Thus, in addition to generating active stress, processive kinesins also increase friction associated with inter-filamentous sliding. Furthermore, while in the dwell state, kinesin motors also act as passive cross-linkers, thus, potentially modifying the elasticity of the active fluids. Another plausible reason for the slowdown at high motor concentrations is the crowding of motors on the MT surface[33, 34].

Negative interference between the K401 motors might decrease the efficiency of interfilament sliding, especially at high motor cluster concentrations. To explore this possibility, we used the non-processive kinesin K365, a monomeric derivative of kinesin-1, which detaches from the MT after each step[35]. Compared to the processive motors, we expect the non-processive K365 to induce less MT cross-linking, and consequently, lower negative interference. Similar to the K401, we formed the clusters of biotinylated K365 with streptavidin (**Fig. 1b**).

Two motors within the same cluster can bind two different MTs, and thus, induce sliding motion. However, clusters can also bind only to a single MT. Such clusters would move along the MT, consuming ATP, but not generating relative filament sliding. In order to reduce the number of such "free-loader" clusters, we permanently attached the kinesin motors to the MT from the neck-end. To accomplish this, we used human kinesin K560 that was labeled with the SNAP-tag, attached to its C-terminal[36, 37-40]. We labelled the MT with Benzyl-Guanine (BG), which covalently reacts with the SNAP-tag labelled kinesin motors (**Fig. 1c**).

The final formulation of an active fluid was based on the passive MT cross-linker, the protein regulator of cytokinesis 1 (PRC1). In this assembly, the ubiquitous depletant was replaced with PRC1, which passively crosslinks the anti-parallel MTs (**Fig. 1d**)[41, 42]. The ubiquitous depletion depends only on the excluded volume interactions. Consequently, most other soft materials dissolved in the active fluid are also adsorbed onto the bundles. This subsequently becomes an obstacle towards making a composite soft active system, consisting of the MTs and the other types of filaments or soft materials; however, PRC1 system overcomes this hurdle.

### III. Materials and Methods

**Kinesin motors and PRC1**: K401-BIO-6xHIS (dimeric MW: 110 kDa) and K365-BIO-6xHIS (monomeric MW: 50 kDa) are the 401 and 365 amino acid N-terminal domains respectively, derived from the Drosophila melanogaster kinesin and fused to the Escherichia coli biotin carboxyl carrier protein, expressed and purified from E. coli[31, 35]. K365-BIO-6xHIS was derived from K401-BIO-6xHIS by deletion mutation using QuikChange lightning multi-site-directed mutagenesis kit (Agilent Technologies), and contains the first 365 amino acids from the drosophila Kinesin domain of the K410-BIO-6xHIS construct. The chains also contain a six-histidine tag, used for affinity purification with a Nickel column. K560-SNAPf-6xHIS (dimeric MW: 168 kDa) is a human conventional kinesin construct, comprised of residues 1–560 with a SNAP-tag and a C-terminal 6-histidine (6xHis) tag[36]. K560-SNAP-6xHIS and K401-BIO-6xHIS are dimeric and processive, whereas K365-BIO-6xHIS is a monomeric and a non-processive kinesin[35, 39]. The Biotin on the K410-BIO-6xHIS and the K365-BIO-6xHIS enables cluster formation through biotin-streptavidin bonding. K560-SNAP-6xHIS is modified to include a SNAP-tag, which is a mutant of O6-alkylguanine-DNA alkyltransferase[39, 40]. SNAP-tag is a DNA repair protein, which binds covalently to BG[37]. All motor proteins were transformed, expressed and purified in Rosetta (DE3) pLysS cells, in accordance with the previously published protocols[43]. The purified proteins were flash frozen in liquid nitrogen with 36% sucrose, and subsequently, stored in -80 °C. The full-length PRC1 (MW: 72.5 kDa) and a truncated form of PRC1 were both transformed and expressed in Rosetta BL21(DE3) cells, and subsequently purified as previously described[41]. The truncated construct, PRC1-NSΔC, consisting of the first 486 amino acids of the full length PRC1 protein, maintains the dimerization, and concurrently, conserves the rod and MT-binding spectrin domains; but, it eliminates most of the unstructured C-terminal domain. The unstructured C-terminal domain has been observed to undergo proteolysis, and is responsible for the interactions between the PRC1 molecules[41]. For the sake of brevity, we refer to K365-BIO-6XHIS, K401-BIO-6XHIS and K560-SNAP-6xHIS as K365, K401, K560-SNAP, PRC1 and PRC1-NSΔC respectively, in the rest of the article.

**Kinesin-streptavidin clusters**: K401 motors dimerize and have two biotin tags, whereas K365 is monomeric with a single biotin tag. K401 and K365 were thawed and incubated with streptavidin (ThermoFisher, 21122, MW: 52.8 kDa) in 1.7:1 biotin to streptavidin ratio, in the presence of DTT. In the case of dimeric kinesin, there was, on average, 0.85 motor per streptavidin molecule. For the K401-streptavidin clusters, 5.7 μL of 6.6 μM streptavidin is mixed with 5 μL of 6.4 μM K401 and 0.5 μL of 5 mM dithiothreitol (DTT) in M2B, and subsequently, left to incubate on ice for 30 minutes. For the K365-streptavidin clusters, 5.7 μL of 6.6 μM streptavidin is mixed with 3.1 μL of 20 μM K365, 0.5 μL of 5 mM DTT and 1.94 μL of an M2B buffer (80 mM PIPES, 1 mM EGTA, 2 mM $MgCl_2$), and then left to incubate on ice for 30 minutes. Henceforth, in this paper, the concentration of motor clusters refers to the concentration of streptavidin in the mixture. These motor clusters were added to the other components of active fluids, and then flash frozen in liquid nitrogen, as discussed later.

**MT polymerization and labelling**: Tubulin (dimeric MW: 100 kDa) was purified from bovine brains, through two cycles of polymerization-depolymerization in high molarity PIPES (1,4-piperazindiethanesulfonic) buffer[44]. Alexa-Fluor 647-NHS (Invitrogen, A-20006) and BG-GLA-NHS (NEB S9151S) labelled tubulins were prepared as previously described[45]. We did not determine the labeling efficiency of BG labelling protocol. The MT polymerization mixture consisted of the unlabelled tubulins, the 3% fluorophore-labeled tubulin, 1 mM DTT, and 0.6 mM GMPCPP (Jena Biosciences, NU-4056) in M2B buffer. The polymerization was done at 37 °C for 30 minutes, at a final tubulin concentration of 80 μM. Subsequently, the polymerization mixture was left to incubate for 6 hours at room temperature, which resulted in MTs of ~1.5 μm length[22]. For the BG-MTs, 6% of the tubulin in the above-mentioned polymerization mixture was BG-labelled. These MTs were added to the other components of active fluids, and then flash frozen in liquid nitrogen, as discussed later.

**Efficiency of K560-SNAP bonding reaction**: Gel electrophoresis was used to determine the efficiency of the K560-SNAP tubulin labeling reaction (**Fig. 2**). Six different amounts of BG-labelled tubulin were reacted with the same amounts of K560-SNAP in the presence of 1 mM DTT. These mixtures were denatured by heating at 100°C, and subsequently loaded for gel electrophoresis. The gel electrophoresis data suggests that even though there are saturating numbers of BG-tubulin, only ~30% of K560-SNAP end up binding to the tubulin, for all the concentrations of tubulin proteins. For the K560-SNAP-based active fluids, a reaction mixture is prepared, consisting of BG-MTs at 44 μM tubulin concentration, appropriate K560-SNAP (ranging from concentrations of 4.23 nM to 47.3 nM), and 1 mM DTT (in Phosphate-buffered saline). The mixture is left to incubate at room temperature for 30 minutes. These K560-SNAP-labeled MTs were added to other components of the active fluids, and then flash frozen in liquid nitrogen.

**Active fluid protocol**: The assembly of active fluids required the mixing of the MTs and a depletion agent with the ATP-

consuming motor clusters. An ATP regeneration system, comprising of phosphoenol pyruvate (PEP, Beantown Chemical, 129745) and pyruvate kinase/lactate dehydrogenase enzymes (PK/LDH, Sigma, P-0294), retained a constant ATP concentration for the duration of the experiment. In addition, we used an oxygen scavenging system, consisting of glucose, DTT, glucose oxidase (Sigma, G2133) and catalase (Sigma, C40), in order to minimize the fluorophore photobleaching.

A pre-mixture was prepared, consisting of 6.66 μL of 0.5 M DTT, 6.66 μL of 300 mg/ml glucose, 6.66 μL of 20 mg/ml glucose oxidase, 6.66 μL of 3.5 mg/ml catalase, 60 μL of 20 mM Trolox (Sigma, 238813), 80 μL of 200 mM PEP, 100 μL of 12 % (w/v) Pluronic (F-127, Sigma P2443. MW: 12.5 kDa), 17 μL of PK/LDH, 29 μl of high-salt buffer (68 mM MgCl2 in M2B), 25 μL of 3 μm sized yellow-green fluorescent polystyrene beads (Polysciences, 18861), and 25 μL of glycerol (functions as a cryoprotectant). 100 μL of freshly polymerized MTs, having 80 μM tubulin concentration, was added to this pre-mixture in a plastic tube. The content of this tube is equally divided in 6 aliquots and the appropriate volumes of the freshly prepared kinesin-streptavidin clusters and M2B were added to these aliquots, resulting in a total volume of 90 μL per aliquot. The content of each aliquot was yet again subdivided into 9 μL samples and flash frozen with liquid nitrogen, and subsequently stored in -80 °C. On the day of the experiment, the frozen sample was thawed, and 1 μL of 14.2 mM of ATP was added to it. This preparation method enhanced the sample reproducibility.

For the K560-SNAP-based active fluids, 6 different BG-MT-K560-SNAP mixtures were prepared. The BG-MT concentrations were kept fixed, whereas the concentrations of the K560-SNAP were varied. Equal volumes of the pre-mixture were added to each BG-MT-K560-SNAP mixture, then aliquoted in 9 μL samples, and finally flash frozen. Reproducibility was tested by preparing three individual samples, all of which were assembled according to these protocols (**Fig. 3b**) For the PRC1 active fluids, 100 μL of freshly polymerized MTs (initial conc. 80 μM) was added to the pre-mixture, containing no Pluronic, then aliquoted in 7.1 μL volumes, and finally flash frozen. On the day of the experiment, the pre-mixture aliquot was thawed, and then 1 μL of ATP of 14.2 mM, and appropriate amounts of PRC1 (or PRC1-NSΔC) and M2B were added.

**Sample chambers:** We used 2 cm×3 mm×100 μm flow channels for our experiments. Glass surfaces were coated with polyacrylamide brush to suppress the surface binding activity of the proteins[46]. To create a flow cell, a spacer is placed between a glass slide (VWR, 25×75×1mm) and a glass coverslip (VWR, 18×18mm, No. 1.5). Spacer material properties influence the sample evolution. Active fluids, made in a channel with a double-sided adhesive tape spacer (Ameritape, 3M9629), decay much faster in comparison to a channel made with a parafilm spacer (Parafilm, PM-996, **Fig. 3a**). To make a flow cell, parafilm was sandwiched between a glass slide and a coverslip, and then heated at 60 °C for a minute. This melted the parafilm, which consequently enabled the cover-slip to stick on the glass slide. Finally, the active fluid was loaded in the channel and sealed with an UV glue.

**Data acquisition and analysis**: To quantify the fluid flows, we doped the fluid with 3 μm passive polystyrene beads, which act as the tracers (**Fig. 4**). The buckling of MT bundles induced large-scale flows; as a result, the tracer particles in the solution moved coherently, with approximately similar speeds[21]. All the data were acquired using the fluorescence microscopy technique. A 4x objective (Plan Fluor, NA 0.13), in conjunction with a CCD camera (Andor, Clara), was used to image the tracer particles, as well as the fluorescently-labeled MTs. To observe the MT bundles, a confocal microscope (Leica TCS SP8) with a 20x objective (Leica HC Fluotar, NA 0.50) was used. The temperature of the sample chambers was consistently maintained at 20 °C. Particle tracking data were acquired at 10 seconds intervals. Lagrangian particle tracking algorithm was used to obtain the trajectories of the tracer particles in the XY-plane[47].

### III. Experimental Results

First, we quantified the temporal behavior of previously studied active fluids, powered by the processive K401 clusters (**Fig. 5a**). We measured the time evolution of the average in-plane speed - denoted as $\langle |v_{xy}| \rangle$ - of the tracer particles. The system self-organized on a timescale that was exclusively dependent on the motor concentration. Upon achieving the maximum activity, fluids did not maintain a constant speed; rather their dynamics slowly started to die down and, after a certain well-defined time, ceased very rapidly. The rapid slowdown might be due to the depletion of the chemical fuel from the regeneration system. We noted that doping the active fluid, after its lifetime, with more ATP regenerating components recovered the activity; consequently, this indicates that the lifetimes of these fluids are dictated by the available energy resources (data not presented here).

The dependence of the maximum speed – denoted as $|v_{max}|$ - on the kinesin cluster concentration was extracted from the speed-time plots (**Fig. 5b**). This data set demonstrates the existence of an optimum motor cluster concentration of 50 nM that maximizes the system dynamics: increasing the motor cluster up to 50 nM speeds up the dynamics, but beyond 50 nM cluster concentration, the active fluid dynamics slows down markedly. Motor cluster concentration also affected other properties of the active fluids. For example, at low concentration of K401 clusters (5 nM) it took up to 3 hours for the system to attain the maximum speed, whereas there was no detectable build-up time at higher motor concentrations. Furthermore, increasing motor concentration resulted in shorter fluid lifetimes (**Fig. 5b**).

We also characterized the structure of the autonomous flows by measuring the spatial velocity-velocity correlation length (denoted as $\frac{\langle v(r).v(r+\Delta r)\rangle}{\langle v(r).v(r)\rangle}$), which decays exponentially as $e^{\frac{-\Delta r}{\lambda}}$. From here, we extracted the correlation length, $\lambda$ (**Fig. 5c**). Below a ~40 μm particle separation distance our observations were unreliable, owing to the ~50 μm depth of focus of the objective (Nikon, 4× Plan Fluor, NA 0.13). The velocity-velocity correlation curves, plotted at different time points during the sample lifetime (data averaged over 1-hour interval), did not collapse on each other; instead they decreased (**Fig. 5d**). Such a decaying trend, in conjunction with the decreasing fluid mean speed, suggests that the structure of the K401 based active fluids changes over time.

Next, we studied active fluids powered by clusters of the non-processive K365 motors (**Fig. 6a**). At 20 nM cluster concentration, the system dynamics slowly increased, attaining its maximum speed after ~3 hours. At higher concentrations, there was no detectable build-up of speed: the optimal self-organized dynamics appeared instantaneously. Once the system attained the maximum speed, there was no substantial drop-off in the speed, especially when compared to the K401 system. The maximum fluid speed increased with increasing K365 cluster concentration; but above the critical value, it plateaued (**Fig. 6b**). This behavior was again different from the K401-based fluids, where the fluid dynamics decreased at cluster concentrations beyond the optimum one. Furthermore, all the correlation curves collapsed on each other, and the correlation lengths remained invariant over the sample lifetimes (**Fig. 6c, d**). This data set indicates that - when compared to the K401 system – the K365 active fluids have increased temporal stability; thus, maintaining more constant dynamics throughout their lifetime. Similar to the K401 system, with increasing K365 cluster concentrations, the fluid lifetimes shorten.

The next system studied was based on the K560-SNAP motors. Most noticeably, the lifetimes of these fluids were considerably longer, when compared to the conventional K401-based fluids at comparable motor concentrations (**Fig. 7**). Similar to other systems, increasing the K560-SNAP concentrations shortened the sample lifetimes (**Fig. 8a**). We suggest that this is connected to the faster ATP consumption rate. The active fluids maintained steady speeds over the entire sample lifetime, for K560-SNAP concentrations of less than 105 nM; however, they exhibited a significant decaying trend at higher motor concentrations. The maximum speed depended non-monotonically on K560-SNAP concentration (**Fig. 8b**). The system dynamics increased up to an optimum concentration, and then slowed down. Furthermore, similar to the K401 system, the velocity correlation function, at different time curves, did not collapse on each other; also, the correlation length $\lambda$ decreased with time (**Fig. 8c, d**).

At high concentrations of SNAP motors, the system exhibited a qualitatively different behavior: the speed dropped drastically with time. To understand this behavior, we have directly imaged the structure of the network with the confocal microscope (**Fig. 9**). At low motor concentrations, the network maintained its structure over the entire sample lifetime. However, at motor concentrations above 105 nM, the network systematically degraded with time. Increasing the Pluronic concentration increased the bundling force, recovering some of the network's stability, even at relatively high motor concentrations. However, the temporal stability in the speed is not recovered by increasing the Pluronic concentration. The mechanism that leads to the bundle disintegration, and the consequent decay in speed, is not easily explicable, since the motors primarily exert forces along the MTs long axis.

In the final active fluid formulation, we replaced the non-specific depletion agent with a MT crosslinker: PRC1; subsequently, we used 250 nM K401 clusters to drive the non-equilibrium dynamics. The average speed of these active fluids had a similar form as the other systems: first, the speed decays slowly, but after a well-defined time, it starts to diminish quite rapidly (**Fig. 10**). We varied the PRC1 concentrations and found that 100 nM optimizes the systems dynamics. For these conditions, the initial dynamics was maximized; however, these systems decayed faster than the active fluids at the other PRC1 concentrations. Decreasing the cross-linked concentration below 100 nM decreased the fluid speed. We suggest that such a decrease is due to the absence of sufficient cross-linkers to bundle all the available MTs. Examining the structure of the active fluids at different PRC1 concentrations with the optical microscope provide support for this hypothesis: the PRC1-based active fluids are much less bundled at 25 nM PRC1, compared to 100 nM and 400 nM PRC1 concentrations respectively. Truncated PRC1 which is discussed later, exhibited the similar behavior (**Fig. 11b**). We also found that the fluid speed decreased beyond the optimal concentration of 100 nM. The decrease in fluid speed beyond the optimum concentration can be explained by a similar argument, as provided previously for other systems: increasing the cross-linker concentration slows down the fluid eventually. We note that the overall sample to sample reproducibility of the PRC1-based fluids was significantly lower, when compared to other systems. This might be related to degradation of the PRC1, which is known to be susceptible to proteolysis[41].

Equivalent experiments were also performed using a truncated version of the full length PRC1, PRC1-NSΔC, which is expected to be more robust to degradation due to the deleted non-structured C-terminus. This data showed behavior qualitatively similar to the full-length protein: 100 nM concentrations of PRC1-NSΔC optimized the maximum speed of the active fluid but were associated with a coarsening bundle structure and the associated decays of mean velocities over time. The magnitude of the average fluid velocities observed in the truncated PRC1 samples were significantly higher than those measured for the full-length protein (**Fig. 11a, b**). Additionally, higher concentrations of PRC1-NSΔC, although relatively slow, display nearly invariant mean velocities; until a sudden collapse ensues, which we expect is due to the depletion of the energy source. Finally, in contrast to the full length PRC1, fluids bundled by PRC1-NSΔC cross-linkers have lifetimes that scale nearly linearly with the concentration of the cross-linker.

## IV. Discussion

We have quantified the temporal stability of K401 fluids, finding that their dynamics systematically drifts with time. In particular, the average speed of the autonomous flows and the structural correlation length significantly decreased over the sample lifetime. Furthermore, increasing the motor concentration beyond a certain threshold led to significant slowdown of the dynamics. We estimate, in Pluronics-bundled K401 active fluids, kinesin step at less than 25-45% the maximum unhindered rate on single microtubules. We calculate this by assuming no futile or backwards steps occur in the kinesin mechanochemical cycle and compare the average hydrolysis of the system (measured by the sample lifetimes) to the maximum hydrolysis rate measured for unhindered kinesin on single microtubules. This can be interpreted in two ways; the actual number of kinesins attached to the MTs may be less than 25-45%. However, this is highly unlikely because the $K_d$ of kinesin is about 37 nM[51]. Another possibility is that the network crosslinked MTs create a stress on each stepping kinesin, limiting its ability to successfully step forward. It is well established that hindering forces decrease the velocity of kinesin and, therefore, their rate of hydrolysis[52]. Indeed, directly crosslinking the MTs with a homogenously crosslinking protein, PRC1-NSΔC is linearly correlated with longer lifetimes. We expect this is not observed in the full length PRC1-based active fluid because this crosslinker is known to behave cooperatively[41]. Potentially, regions of densely crosslinked MTs dominate the stresses kinesin must overcome throughout the range of examined PRC1 concentrations.

These findings present a potential difficulty for quantitative comparison to theoretical models. We hypothesize that these changes are due to the negative interference between the processive K401 motors, as discussed previously[32]. To test this hypothesis, we studied active fluids powered by the non-processive single-headed K365 motors, finding that these systems exhibited superior properties. The average speed and the spatial correlation length of K365 fluids remained fairly constant over the entire sample lifetime. Furthermore, unlike the K401 system, the K365 active fluids did not exhibit a decrease in the system speed at high clusters concentrations. We suggest that this is due to the reduced interference between the non-processive K365 motors. In particular, it will be useful to quantify the dynamics of the active nematic liquid crystals powered by K365 motors, as the recent experiments have suggested that, at low ATP concentrations, the conventional K401 motors might significantly modify the liquid crystal elastic constants[48].

Force generation by the single headed kinesin motors have been studied using motility assays[35, 49, 50]. Compared to this work, the observation that the K365 clusters efficiently generate interfilament sliding raises several important questions. First, we found that the speed of the K365 fluids is comparable to the conventional K401 systems. This is in contrast to the motility experiments, where the single-headed kinesin-generated velocities were significantly smaller than those of the double-headed motors. Second, in the motility assays single-headed motors generated motility only when multiple motors were interacting with the MT simultaneously. A minimum of 4 to 6 motors were required to generate any motion. We observed significant dynamics at cluster concentrations as low as 20 nM. Under these conditions, there are only about four motor clusters per MT, similar to the number of single-headed motors that need to be engaged with a MT to generate motion in a motility assay. However, in a motility assay the motors are permanently attached to the solid surface: each step imparts force on the sliding MT. In active fluids, clusters need to be engaged with both the MTs to generate sliding motion. However, the low duty ratio of single-headed motors reduces the probability of such occurrences. These inconsistencies suggest insufficient understanding of how the single-headed motors generate MT sliding.

We showed that the lifetime of active fluids can be extended to multiple days by linking the motor proteins to the MT backbone. We suggest that the longevity of these samples stems from eliminating the population of "free-loader" kinesin clusters that bind to MTs by a single motor, thus consuming chemical fuel, but failing to generate any active stress. These long-lived active fluids should enable characterization of the non-equilibrium phenomena, where repeated time averaging is required to obtain quantitative data. For example, they might shed light on the nature of the transition from the chaotic to the coherent flows that is observed in confined active isotropic fluids[20], as well as the dynamical states that emerge in the assemblages of motile active emulsions[19].

In all the previously studied active fluids, the MTs were bundled with the depletant polymers. Depletion forces are non-specific among various soft materials, thus greatly limiting control over the structures of the bundled MT active fluids, combined with other soft materials. We introduced PRC1 as a cross-linking agent specific to the anti-parallel MTs. PRC1-based active fluids open the door for assembling systems, consisting of various types of soft materials, alongside MTs. Given PRC1's specific preference for the MTs, the secondary material will remain immune to the bundling process. The PRC1-based active fluids enable a method for assembling composite systems, wherein the dilute MT fluid generates active stresses, which subsequently drive the passive component away from equilibrium.

In summary, we studied active fluids that were assembled from distinct microscopic units. All the fluids exhibited similar large-scale dynamics, which was powered by the mesoscopic extensile MT bundles. This suggests that the dynamics of the active isotropic fluids is highly robust against variations at the microscopic level. The knowledge generated from our studies should be easily transferrable to other manifestations of MT-based active matter, such as active nematics.

**Conflicts of interests**: There are no conflicts to declare.

**Acknowledgements:** This study was primarily supported by the Department of Energy of Basic Energy Sciences, through award DE-SC0010432TDD. Design of the proteins for this work were supported by Brandeis MRSEC, through grant NSF-MRSEC-1420382. We also acknowledge the use of a MRSEC optical and biosynthesis facility supported by NSF-MRSEC-1420382. The K560-SNAP used in this work was a gift from S. L. Reck-Peterson.

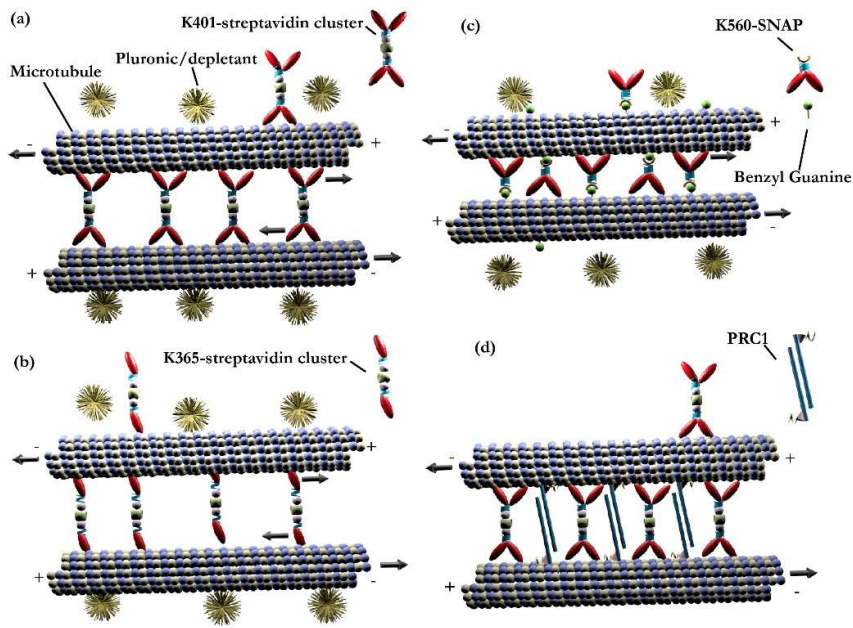

**Fig. 1. Microscopic building blocks of four active isotropic fluids. (a)** Previously studied active fluids consisted of MTs, clusters of double-headed processive kinesin motors (K401) and a depleting agent. **(b)** Schematic of active fluid powered by clusters of single-headed non-processive kinesin motors (K365). **(c)** Active fluid powered by double-headed processive kinesin K560-SNAP that is covalently bonded to the MT. **(d)** Active fluid in which non-specific depletion agent was replaced by PRC1, a protein that specifically crosslinks anti-parallel MTs.

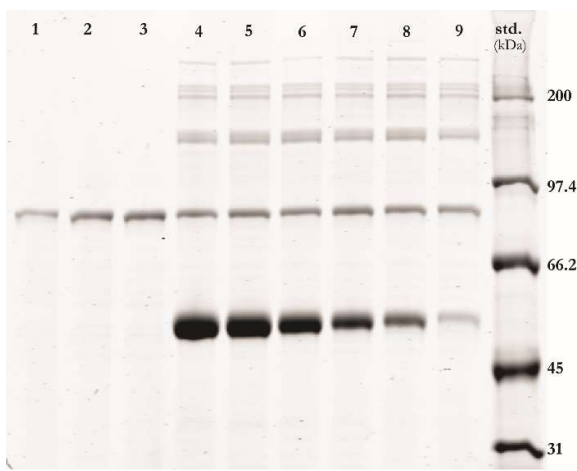

**Fig. 2. Efficiency of the SNAP based bonding reaction estimated by gel electrophoresis** Lane 1-3 contain 0.87, 1.31, 1.75 µg of K560-SNAP, respectively. Lane 4-9 contains same amount of K560-SNAP incubated with decreasing amounts of BG tubulins (5, 3.75, 2.5, 1.87, 1.25 and 0.62 µg of BG-tubulins). Lane 10 contains the standard, SDS-broad range protein (Bio-Rad, 161-0317). The bands at 50 kDa are unlabelled tubulin monomer; the bands at 84 kDa correspond to unreacted K560-SNAP, and the band at ~134 kDa correspond to linked K560-SNAP-BG-tubulin complex. All the bands at and above 200 kDa correspond to multimers of K560-SNAP attached to same BG-tubulin, which can take place presumably due to two SNAP tags associated with each K560-SNAP. Using the intensity versus K560-SNAP amount plot obtained from lane 1-3, we determined the amount of unreacted K560-SNAP in lanes 4-9.

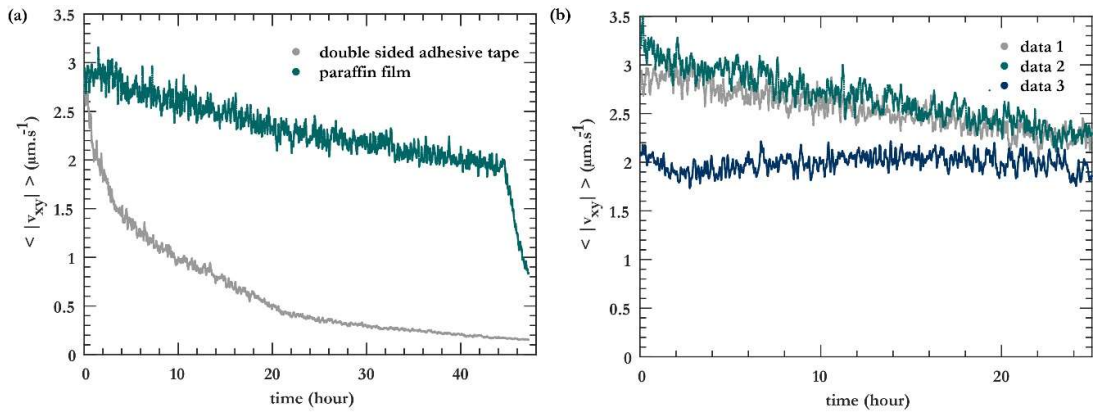

**Fig. 3. Sensitivity to type of confinement and reproducibility of the active fluids**. **(a)** Dynamics of the active fluid is sensitive to the chemical nature of the flow chamber. The flow chamber prepared with adhesive tape leads to sample degradation, presumably due to leaching of chemical. **(b)** Typical reproducibility of the active fluid dynamics with three samples that are nominally the same but prepared independently of each other.

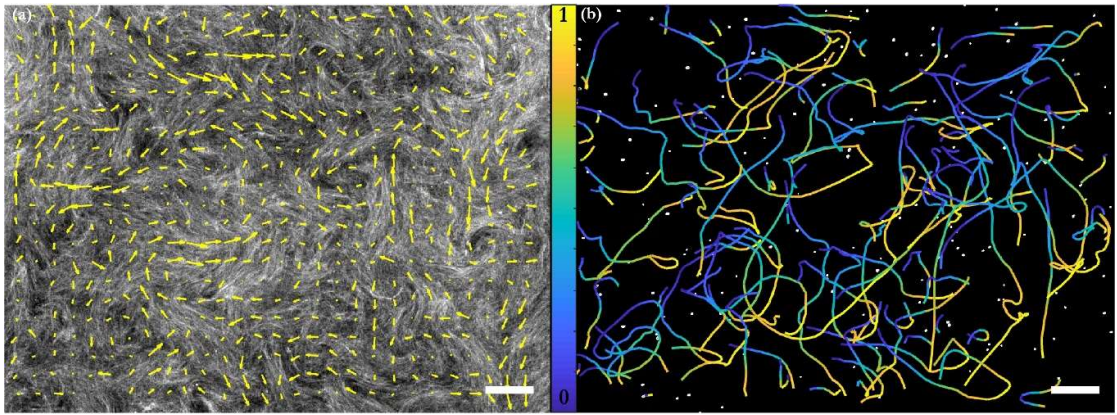

**Fig. 4. Large scale flows of isotropic active fluid.** **(a)** The instantaneous flow field of active isotropic fluid is quantified by visualizing fluorescently labelled MTs; yellow arrows represent the velocity field obtained using PIV. Scale bar is 200 μm. **(b)** Trajectories of passive tracer particles (bright dots) embedded in and advected by the active fluid. Colour bar represents the passage of time from 0 to 1 minute, scale bar is 200 μm.

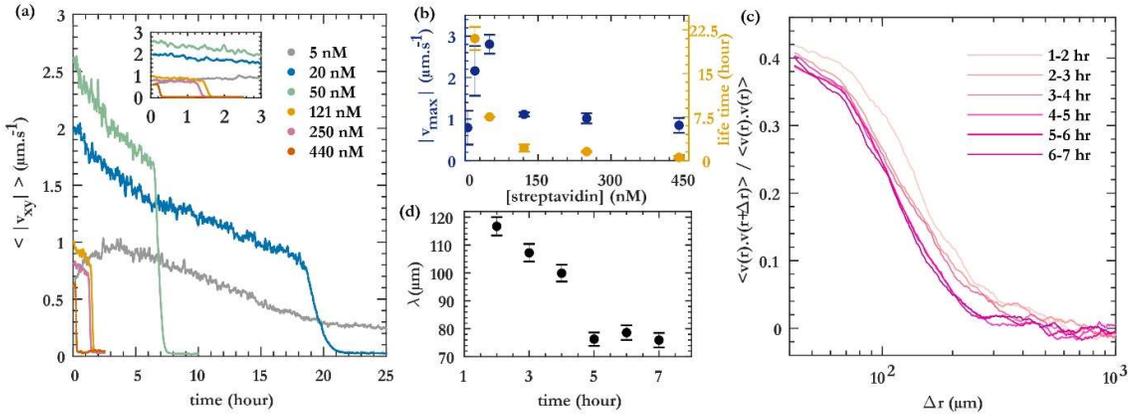

**Fig. 5. Dynamics of active fluids powered by processive K401 motors. (a)** Time dependence of the mean speed of tracer particles advected by the active fluid. Data is shown for different concentration of K401 clusters. **(b)** Maximum speed of tracer paticles and the sample lifetime as a function of K401 cluster concentration. **(c)** Spatial velocity-velocity correlation of tracer particles plotted as a function of particle separation, for 50 nM cluster concentration. The spatial velocity correlation is average over one hour. **(d)** Time evolution of the correlation length, λ, extracted from the spatial correlation function.

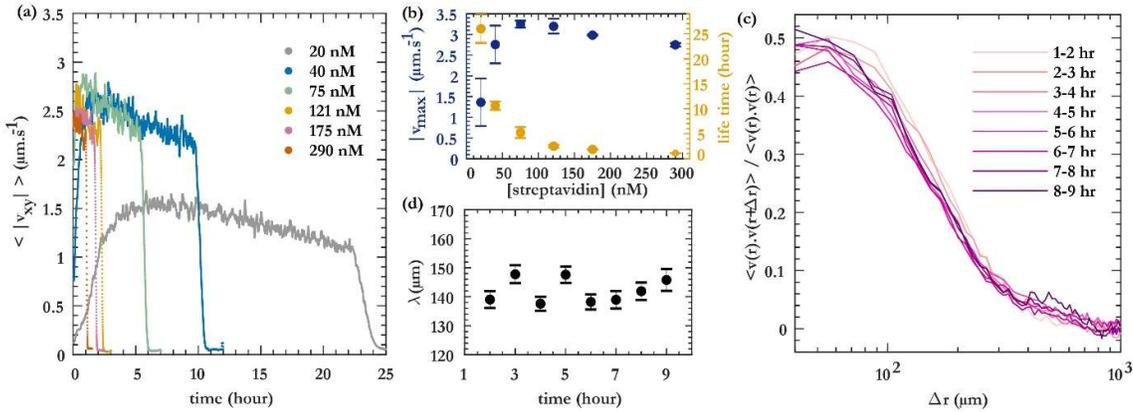

**Fig. 6. Dynamics and structure of active fluid powered by non-processive K365 motors. (a)** Time dependence of the mean speed of tracer particles advected by the active fluid. Data is shown for different concentration of K365 clusters. **(b)** Maximum speed of tracer paticles and the sample lifetime as a function of K365 cluster concentration. **(c)** Spatial velocity-velocity correlation of tracer particles plotted as a function of particle separation, for 50 nM cluster concentration. The spatial velocity correlation is average over one hour. **(d)** Time evolution of the correlation length, λ, extracted from the spatial correlation function.

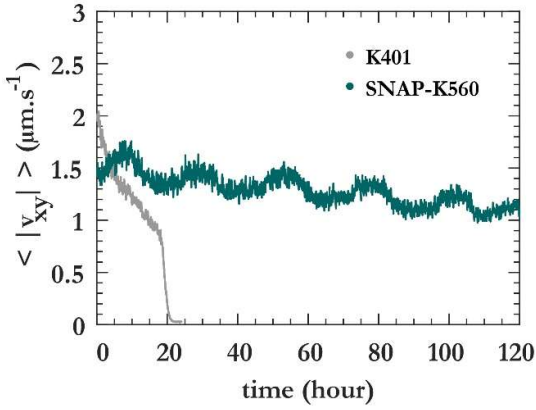

**Fig. 7. Extending the active fluid lifetime by linking motors to MTs.** Mean speed of tracer particles, measured over time, for K401 and K560-SNAP based active fluids. The K560-SNAP sample was not temperature controlled and the 24-hour oscillations are caused by diurnal temperature fluctuations of the building.

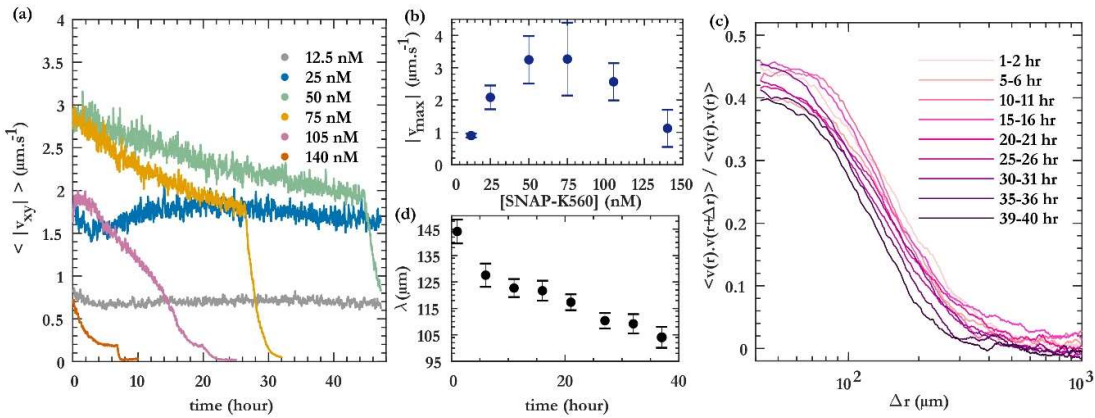

**Fig. 8. Dynamics and structure of active fluid powered by K560-SNAP motors linked to MT. (a)** Time dependence of the mean speed of tracer particles advected by the active fluid. Data is shown for different concentration of K560-SNAP clusters. **(b)** Maximum speed of tracer paticles and the sample lifetime as a function of K560-SNAP cluster concentration. **(c)** Time evolution of the velocity-velocity correlation of tracer particles plotted as a function of particle separation, for 50 nM cluster concentration. The spatial velocity correlation is average over one hour. **(d)** Time evolution of the correlation length, $\lambda$, extracted from the spatial correlation function.

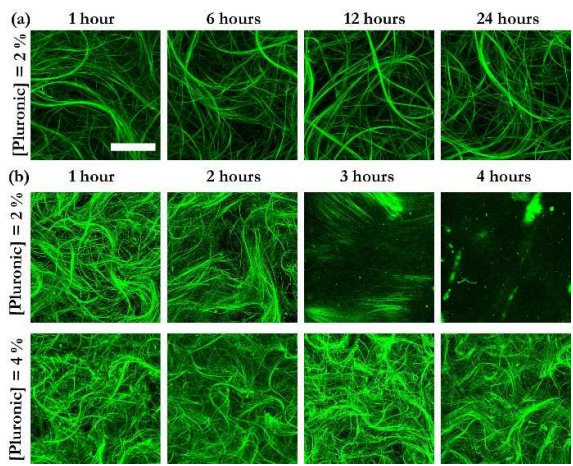

**Fig. 9. Structure of K560-SNAP based active fluids at different depletant and motor concentrations. (a)** Structure of active fluid consisting of 25 nM K560-SNAP, observed over time with confocal microscopy. **(b)** Structure of the active fluid consisting 140 nM K560-SNAP, observed over time, at two different Pluronic (depletant) concentrations. Scale bar, 100 μm.

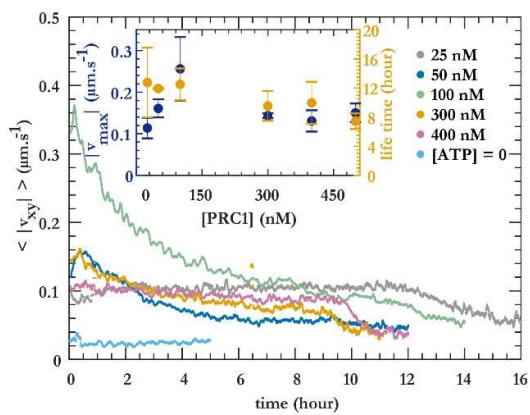

**Fig. 10. Dynamics of PRC1-based active fluids.** Dependence of mean speed of tracer particles on the PRC1 concentrations. Inset : maximum speed of tracer particles and the life time of the active fluid as a function of PRC1 concentration.

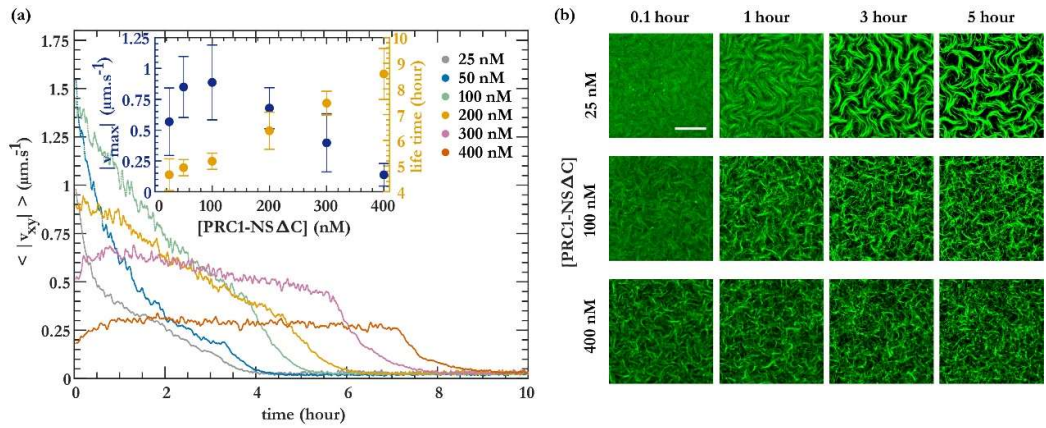

**Fig. 11. Enhanced dynamics with truncated PRC1. (a)** Dependence of mean speed of tracer particles on the PRC1-NSΔC concentrations. Inset : maximum speed of tracer particles and the life time of the active fluid as a function of PRC1-NSΔC concentration. **(b)** Structure of active fluids consisting of different amounts of PRC1-NSΔC observed with fluorescence microscopy. Scale bar, 500 μm.